\documentclass[12pt]{article}

\usepackage{latexsym} 
\usepackage{amssymb}  
\usepackage{amsfonts} 
\usepackage{amsbsy}   
\usepackage{amsmath} 
\usepackage{graphicx}       
\usepackage[dvips]{color}
\usepackage{bm}   

\newcommand{\txt}{\textstyle}

\newcommand{\third}{{\txt \frac{1}{3}}}

\makeatletter 

\def\appendix{\par                              
    \setcounter{section}{0}                     
    \setcounter{subsection}{0}
    \renewcommand{\theequation}{\Alph{section}.\arabic{equation}}
    \renewcommand{\thesection}{Appendix \Alph{section}
                \setcounter{equation}{0}  } 
    \renewcommand{\thesubsection}{\Alph{section}.\arabic{subsection}}
}

\def\applabel#1{\@bsphack
  \protected@write\@auxout{}%
         {\string\newlabel{#1}{{\Alph{section}}{\thepage}}}%
  \@esphack}

\def\section{
\setcounter{equation}{0}        
\@startsection {section}{1}{\z@}{-3.5ex plus -1ex minus 
 -.2ex}{2.3ex plus .2ex}{\large\bf}}
\renewcommand{\theequation}{\arabic{section}.\arabic{equation}}

\def\subsection{\@startsection{subsection}{2}{\z@}{-3.25ex plus -1ex minus 
 -.2ex}{1.5ex plus .2ex}{\normalsize\bf}}

\def\subsubsection{\@startsection{subsubsection}{3}{\z@}{-3.25ex plus
 -1ex minus -.2ex}{1.5ex plus .2ex}{\normalsize}}

\makeatother   

\begin{document}

 
\title{\bf Level Crossings in Complex Two-Dimensional Potentials}

\author{Qing-hai Wang \\[2ex]
Department of Physics\\
National University of Singapore\\ 
Singapore 117542}

\date{April 3, 2009}

\begin{titlepage}
\maketitle
\renewcommand{\thepage}{}          

\begin{abstract}
Two-dimensional ${\cal PT}$-symmetric quantum-mechanical systems with the complex cubic potential $V_{12}=x^2+y^2+igxy^2$ and the complex H\'enon-Heiles potential $V_{\rm HH}=x^2+y^2+ig\left(xy^2-x^3/3\right)$ are investigated. Using numerical and perturbative methods, energy spectra are obtained to high levels. Although both potentials respect the ${\cal PT}$ symmetry, the complex energy eigenvalues appear when level crossing happens between same parity eigenstates.
\end{abstract}

{\bf PAC numbers: 03.65.Ge, 02.30.Lt, 02.60.Lj}
\end{titlepage}

\section{Introduction}

In 1998, Bender and Boettcher introduced the non-Hermitian ${\cal PT}$-symmetric quantum systems \cite{BB}. This letter ignited a very active field. Hundreds of papers have been published on various aspects of this new quantum mechanics. For review, please see Ref.~\cite{Review} and references therein.

Most of the studies have been focused on one-dimensional systems, very few have touched two or higher dimensions. In 2001, Bender {\it et al}\ studied the complex cubic potential $V_{12}=x^2+y^2+igxy^2$ 
and $V_{111}=x^2+y^2+z^2+igxyz$. Using perturbation theory, they found real eigenvalues for the three lowest levels. They used the WKB method to confirm the reality of the ground state energy. Their approach also applied to the complex H\'enon-Heiles potential $V_{\rm HH}=x^2+y^2+ig\left(xy^2-x^3/3\right)$ \cite{BDMS}. In 2002, Nanayakkara and Abayaratne revisited the $igxy^2$ interaction with a non-degenerated setup. They computed energy eigenvalues up to $n=4$. With their choice of parameters, all eigenvalues appear to be real \cite{NanaPLA1}. Later in the same year, Nanayakkara obtained the analytic results for $igxy$ interaction and numerical results in $igxy^3$ and $igxyz^2$ interactions. For all three systems, the author found real spectrum for small enough coupling constant and complex one for large coupling. The interaction of $igx^2y^2$ is also studied \cite{NanaPLA2}. In 2005, Nanayakkara observed avoided level crossings in both the real cubic interaction $gxy^2$ and the complex cubic interaction $igxy^2$ \cite{NanaPLA3}. The results has been challenged in a comment by B\'ila {\it et al}\ in Ref.~\cite{BTZPLA}. They pointed out that the quantum system with the real cubic potential is ill defined and the complex cubic potential has no avoided level crossings. 

In this paper, we compute the eigenvalues of two-dimensional systems to high levels with high precision. We study the Hamiltonians associated with the complex cubic potential $V_{12}$ 
\begin{eqnarray}
H_{12} &=& p_x^2 + p_y^2 + x^2 + y^2 + igxy^2
\label{eqn:H12}
\end{eqnarray}
and with the complex H\'enon-Heiles potential $V_{\rm HH}$
\begin{eqnarray}
H_{\rm HH} &=& p_x^2 + p_y^2 + x^2 + y^2 + ig\left( xy^2-\third x^3\right).
\label{eqn:Hhh}
\end{eqnarray}
Both Hamiltonians are not Hermitian. Rather, they respect the ${\cal PT}$ symmetry and the $y$ parity:
\begin{eqnarray}
[H,{\cal PT}]=0,&\qquad& [H,{\cal P}_y]=0,
\end{eqnarray} 
where the total parity ${\cal P}$, the time reversal ${\cal T}$, and the $y$ parity ${\cal P}_y$ are defined as
\begin{eqnarray}
{\cal P}&:& i\to i;\quad x\to -x;\quad y\to -y;\nonumber \\
{\cal T}&:& i\to -i;\quad x\to x;\quad y\to y; \\
{\cal P}_y&:& i\to i;\quad x\to x;\quad y\to -y. \nonumber
\end{eqnarray} 
Although the Hamiltonians are complex, the secular equations are real due to the ${\cal PT}$ symmetry. They can only depend on $g^2$. Therefore, both systems are invariant under $g\to -g$. 

Since the $y$ parity is an unbroken symmetry, we can diagonalize $H$ and ${\cal P}_y$ simultaneously. All the eigenfunctions of $H$ are also eigenfunctions of ${\cal P}_y$ with eigenvalues $\pm 1$. 

We use different methods to compute the eigenvalues of each system. We found complex conjugate pairs when the levels cross between same $y$-parity states. At higher level, the level crossing happens at smaller coupling constant. This indicates that no matter how small the coupling constant is, there are always complex eigenvalues at high enough levels. It shows the rich structure in the two-dimensional ${\cal PT}$-symmetric quantum systems. One must be very careful when generalizing the results in one-dimensional non-Hermitian quantum mechanics to higher dimensions.

The paper is organized as the following: the methods we used are briefly reviewed in Sec.~\ref{sec:method}; the results of different systems are presented in Sec.~\ref{sec:xy2} and Sec.~\ref{sec:HH}; we conclude in Sec.~\ref{sec:conclusion}.

\section{Methodology}
\label{sec:method}

We use three different methods to study the spectra of Hamiltonians in (\ref{eqn:H12}) and (\ref{eqn:Hhh}). 

\begin{itemize}
\item The first method is the perturbation theory. Because the energy levels of unperturbed Hamiltonian, 
$$H_0=p_x^2+p_y^2+x^2+y^2,$$ 
has degeneracy, we must apply the degenerate perturbation theory. After getting the eigenvalues as power series of the coupling constant $g$, we use Pad\'e expansions to extract the information from the divergent series. The results for two systems are shown as lines in Fig.~\ref{fig:xy2_Re} and Fig.~\ref{fig:HH_Re}, respectively.
\item The second method is numerical computation by the finite-element method (FEM). The FEM results for eigenvalues with $V_{12}$ potential are shown as crosses in Fig.~\ref{fig:xy2_Re} and Fig.~\ref{fig:xy2_Im}. 
\item The third method is based on the expansion in two-dimensional harmonic oscillator (HO) basis \cite{Mathematica}. We first analytically compute the non-vanishing matrix elements of the full Hamiltonian on the two-dimensional HO eigenfunctions basis. Depending on the precision requirement, we truncate the sparse matrix to a finite size. And the eigenvalues of the full Hamiltonian can be obtained by numerically diagonalizing this finite matrix. This method is a mixture of analytic and numerical techniques. It turns out that this is the best way to compute the eigenvalues of the two-dimensional complex Hamiltonians. The results are shown as dots in Fig.~\ref{fig:xy2_Re} and Fig.~\ref{fig:xy2_Im} and crosses in Fig.~\ref{fig:HH_Re} and Fig.~\ref{fig:HH_Im}.
\end{itemize}

From Fig.~\ref{fig:xy2_Re} we can clearly to see that all three methods provide consistent informations about the real part of the eigenvalues. And Fig.~\ref{fig:xy2_Im} shows the two numerical methods are consistent on the imaginary part of the eigenvalues.

\section{Eigenvalues with the complex cubic potential}
\label{sec:xy2}

\begin{figure}[htbp]
\begin{center}
\includegraphics[width=0.75\textwidth]{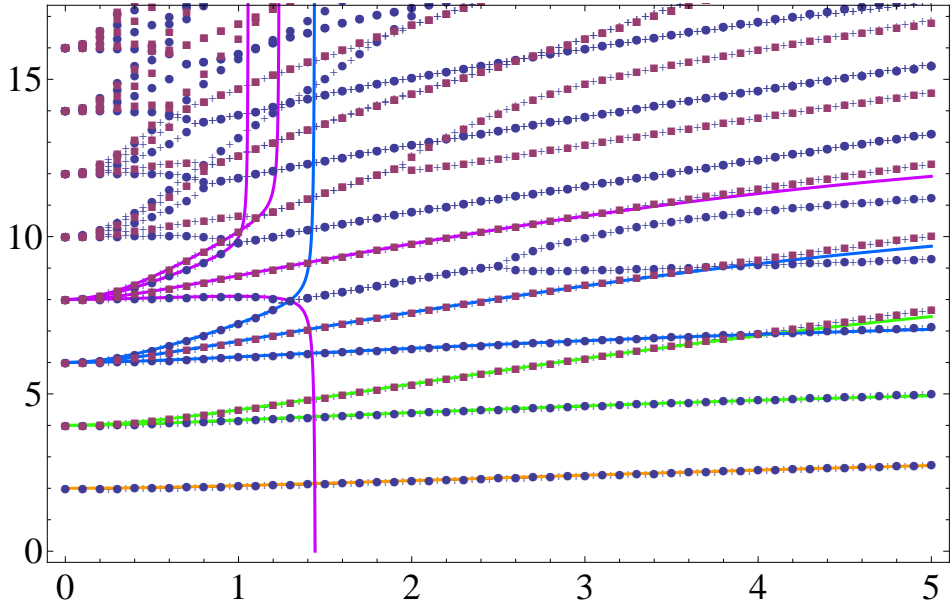}
\end{center}
\caption{\small
Real parts of the eigenvalues as functions of the coupling constant $g$ with the complex cubic potential. Lines are from perturbative expansion using $(20,20)$ Pad\'e. Crosses are results using FEM. Dots are results using the method based on two-dimensional HO basis expansions. Blue disks have even $y$-parity and purple squares have odd $y$-parity. The system is symmetric under $g\to -g$, only the part with positive $g$ is shown.}
\label{fig:xy2_Re}
\end{figure}

Using perturbation theory, we calculate the eigenvalues as power series of the coupling constant $g$. Here are the first few terms for the first 10 eigenvalues:
\begin{eqnarray}
E_{00} &=& 2 + \frac{5}{48}g^2 - \frac{223}{6912}g^4 + \frac{114407}{4976640}g^6 - \frac{346266143}{14332723200}g^8 +\cdots\nonumber \\
E_{10}&=& 4 + \frac{11}{16}g^2 - \frac{869}{2304}g^4 + \frac{737419}{1658880}g^6 - \frac{3486539861}{4777574400}g^8 +\cdots\nonumber \\
E_{11}&=& 4 + \frac{13}{48}g^2 - \frac{1519}{6912}g^4 + \frac{1535767}{4976640}g^6 - \frac{7858558079}{14332723200}g^8 +\cdots\nonumber \\
E_{20}&=& 6 + \left(\frac{17}{16}+\frac{\sqrt{41}}{8}\right)g^2 - \left(\frac{329}{384} + \frac{3407}{768\sqrt{41}}\right)g^4 + \left(\frac{417793}{276480}\right. \nonumber \\
&&\qquad \left. + \frac{63502133}{7557120\sqrt{41}}\right)g^6 - \left(\frac{952249153}{265420800} + \frac{18548037835009}{892344729600 \sqrt{41}}\right)g^8 +\cdots\nonumber \\
E_{21}&=& 6 + \frac{19}{16}g^2 - \frac{1063}{768}g^4 + \frac{1606697}{552960}g^6 - \frac{4024837709}{530841600}g^8 +\cdots\nonumber \\
E_{22}&=& 6 + \left(\frac{17}{16}-\frac{\sqrt{41}}{8}\right)g^2 - \left(\frac{329}{384}-\frac{3407}{768\sqrt{41}}\right)g^4 + \left(\frac{417793}{276480}\right. \nonumber \\
&&\qquad \left.-\frac{63502133}{7557120\sqrt{41}}\right)g^6 - \left(\frac{952249153}{265420800} - \frac{18548037835009}{892344729600\sqrt{41}}\right)g^8 +\cdots\nonumber \\
E_{30}&=& 8 + \left(\frac{115}{48}+\frac{\sqrt{721}}{24}\right)g^2 - \left(\frac{9205}{3456} + \frac{260275}{6912\sqrt{721}}\right)g^4 + \left(\frac{3128263}{497664}\right. \nonumber \\
&&\qquad \left. +\frac{77128555369}{717631488\sqrt{721}}\right)g^6 - \left(\frac{28693057087}{1433272320} + \frac{576524526420731587}{1490147432202240\sqrt{721}}\right)g^8 +\cdots\nonumber \\
E_{31}&=& 8 + \frac{137}{48}g^2 - \frac{888811}{214272}g^4 + \frac{1766794711427}{148259082240}g^6 - \frac{16887386781611073971}{410333696734003200}g^8 +\cdots\nonumber \\
E_{32}&=& 8 + \left(\frac{115}{48}-\frac{\sqrt{721}}{24}\right)g^2 - \left(\frac{9205}{3456} - \frac{260275}{6912\sqrt{721}}\right)g^4 + \left(\frac{3128263}{497664}\right. \nonumber \\
&&\qquad \left. -\frac{77128555369}{717631488\sqrt{721}}\right)g^6 - \left(\frac{28693057087}{1433272320} - \frac{576524526420731587}{1490147432202240\sqrt{721}}\right)g^8 +\cdots\nonumber \\
E_{33}&=& 8 + \frac{13}{48}g^2 - \frac{85457}{214272}g^4 + \frac{126990201721}{148259082240}g^6 - \frac{998074124043859297}{410333696734003200}g^8 +\cdots
\end{eqnarray}
We actually continue this expansion up to the order of $g^{40}$, then use $(20,20)$ Pad\'e to extract the information from the divergent series. The results are shown as lines in Fig.~\ref{fig:xy2_Re}. From the figure, we may see that some Pad\'e expansions have poles and cross each others. 

In the same figure, we show the real parts of eigenvalues from FEM as crosses and from HO expansions as dots. The two sets of numerical results clearly agree with each other. And surprisingly, up to the crossing points, the Pad\'e expansions fit the numerical results very well.  

\begin{figure}[htbp]
\begin{center}
\includegraphics[width=0.75\textwidth]{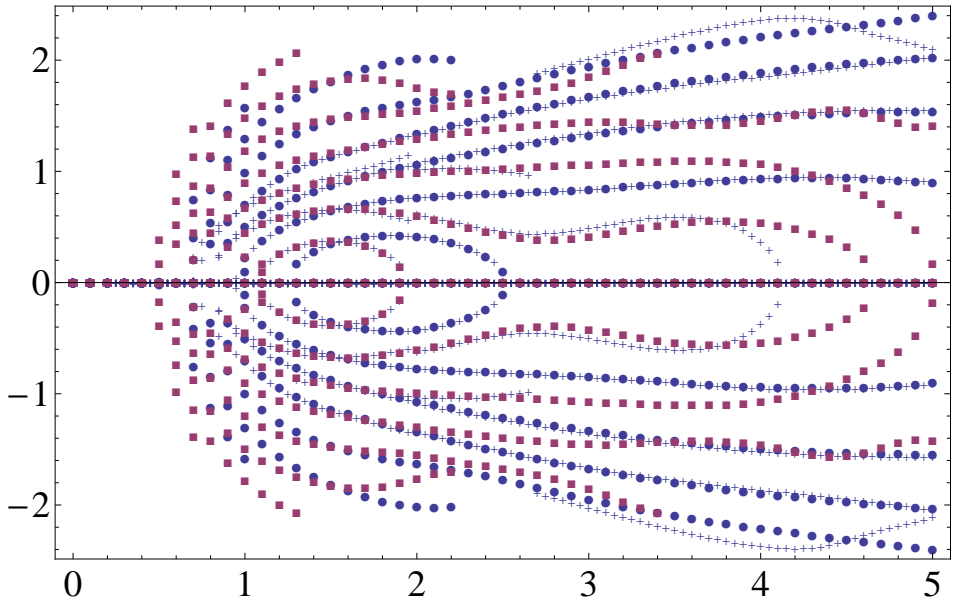}
\end{center}
\caption{\small
Imaginary parts of the eigenvalues as functions of the coupling constant $g$ with the complex cubic potential. Crosses are results using FEM. Dots are results using the method based on two-dimensional HO basis expansions. Blue disks have even $y$-parity and purple squares have odd $y$-parity. The system is symmetric under $g\to -g$, only the part with positive $g$ is shown.}
\label{fig:xy2_Im}
\end{figure}

Although the Hamiltonian $H_{12}$ has ${\cal PT}$ symmetry, some energy eigenvalues are complex. The imaginary parts of the eigenvalues by two numerical methods are shown in Fig.~\ref{fig:xy2_Im}.

From the Fig.~\ref{fig:xy2_Re}, we observe two types of level crossings. If the crossing is between different $y$-parity states, the eigenvalues remain real. One example of this type is the crossing between the third and fourth lowest states near $g=4$. If the crossing is between same $y$-parity states, then eigenvalues become complex conjugate pairs. For example, the sixth and seventh lowest states cross near $g=1.4$. 

We also observed that the level crossing happens in higher levels at smaller value of coupling constant. This indicates that no matter how small the coupling constant is, there is always complex eigenvalues at high enough levels. There is no critical value, $g_c>0$, such that for $|g|<g_c$, the entire spectrum is real. 

\section{Eigenvalues with the complex H\'enon-Heiles potential}
\label{sec:HH}

In the case of the complex H\'enon-Heiles potential, a similar patten appears. The energy eigenvalues become complex when $g\neq 0$. There are special features for this interaction. 
\begin{itemize}
\item First, there are all rational numbers in every term of eigenvalues in the perturbation expansion. 
\item Second, there are an accident degeneracy between odd and even $y$-parity. Most of the energy levels are degenerated, but interestingly, not all of them. For example, $E_{10}=E_{11}$, $E_{21}=E_{22}$, and $E_{30}=E_{31}$, but $E_{32}\neq E_{33}$. In fact, $E_{32}$ and $E_{33}$ are the same up to the order of $g^2$, but for the order of $g^4$ and higher, they separate. This patten remains the same for higher exciting states:
\begin{itemize}
\item For even $n\geq 2$, $E_{n1}= E_{n2}$, $E_{n3}= E_{n4}$, etc.
\item For odd  $n\geq 3$, $E_{n0}= E_{n1}$, $E_{n4}= E_{n5}$, etc.
\item For odd  $n\geq 3$, $E_{n2}\neq E_{n3}$ starting from the order of $g^4$.
\end{itemize} 
\item Third, $E_{53}$ appears to be non-alternating in signs in the expansion. And this is a unique case up to $n=9$.
\end{itemize}

\begin{figure}[htbp]
\begin{center}
\includegraphics[width=0.75\textwidth]{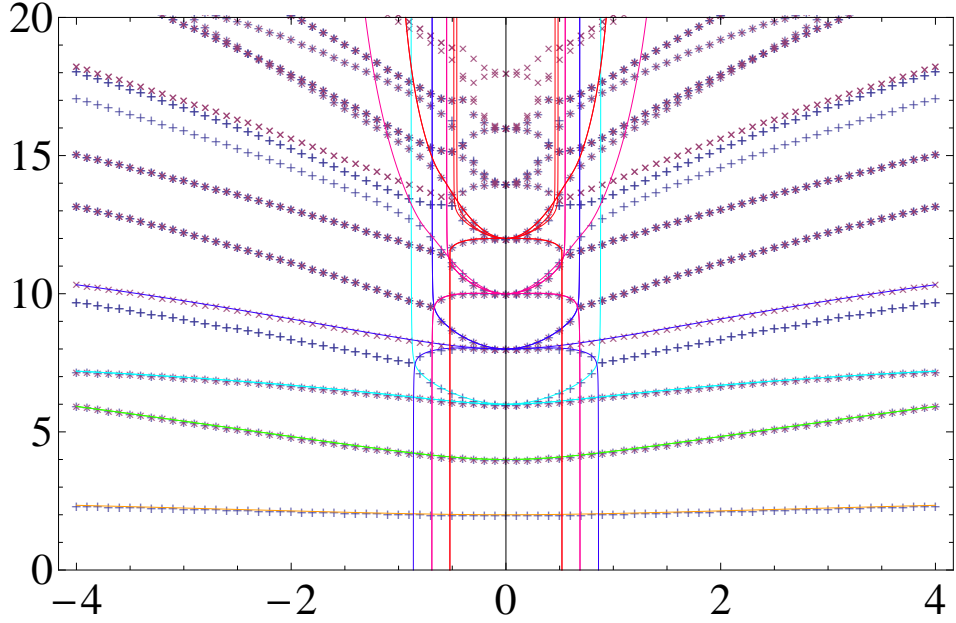}
\end{center}
\caption{\small
Real parts of the eigenvalues as functions of the coupling constant $g$ with the complex H\'enon-Heiles potential. Lines are from perturbative expansion using $(20,20)$ Pad\'e. Crosses are from HO expansions. Even $y$-parity states are using ``$+$'' and odd $y$-parity states are using ``$\times$''.}
\label{fig:HH_Re}
\end{figure}

Here are the first few terms for the first 21 eigenvalues:
\begin{eqnarray}
E_{00} &=& 2 + \frac{1}{18}g^2 - \frac{11}{864}g^4 + \frac{6089}{933120}g^6 - \frac{2221951}{447897600}g^8 +\cdots\nonumber \\
E_{10}=E_{11}&=& 4 + \frac{7}{18}g^2 - \frac{133}{864}g^4 + \frac{30191}{233280}g^6 - \frac{67779467}{447897600}g^8 +\cdots\nonumber \\
E_{20}&=& 6  + \frac{31}{18}g^2 - \frac{145}{288}g^4 + \frac{200923}{186624}g^6 - \frac{40752209}{29859840}g^8 +\cdots\nonumber \\
E_{21}=E_{22}&=& 6  + \frac{5}{9}g^2 - \frac{83}{144}g^4 + \frac{432493}{466560}g^6 - \frac{133188257}{74649600}g^8 +\cdots\nonumber \\
E_{30}=E_{31}&=& 8 + \frac{26}{9}g^2 - \frac{535}{432}g^4 + \frac{180037}{46656}g^6 - \frac{296084959}{44789760}g^8 +\cdots\nonumber \\
E_{32}&=& 8  + \frac{5}{9}g^2 - \frac{1123}{432}g^4 + \frac{1416869}{233280}g^6 - \frac{3963323843}{223948800}g^8 +\cdots\nonumber \\
E_{33}&=& 8  + \frac{5}{9}g^2 - \frac{115}{432}g^4 + \frac{12121}{46656}g^6 - \frac{15676999}{44789760}g^8 +\cdots\nonumber \\
E_{40}&=& 10 + \frac{91}{18}g^2 - \frac{2065}{864}g^4 + \frac{1208431}{186624}g^6 - \frac{1731827209}{89579520}g^8 +\cdots\nonumber \\
E_{41}=E_{42}&=& 10  + \frac{35}{9}g^2 - \frac{1085}{432}g^4 + \frac{1285823}{93312}g^6 - \frac{1478364167}{44789760}g^8 +\cdots\nonumber \\
E_{43}=E_{44}&=& 10  + \frac{7}{18}g^2 - \frac{2485}{864}g^4 + \frac{1063615}{186624}g^6 - \frac{1819581169}{89579520}g^8 +\cdots\nonumber \\
E_{50}=E_{51}&=& 12  + \frac{127}{18}g^2 - \frac{1205}{288}g^4 + \frac{814129}{46656}g^6 - \frac{1958220799}{29859840}g^8 +\cdots\nonumber \\
E_{52}&=& 12  + \frac{85}{18}g^2 - \frac{2633}{288}g^4 + \frac{1370563}{29160}g^6 - \frac{20818356203}{149299200}g^8 +\cdots\nonumber \\
E_{53}&=& 12  + \frac{85}{18}g^2 + \frac{55}{288}g^4 + \frac{70673}{5832}g^6 + \frac{354058961}{29859840}g^8 +\cdots\nonumber \\
E_{54}=E_{55}&=& 12  + \frac{1}{18}g^2 - \frac{1457}{288}g^4 + \frac{329257}{29160}g^6 - \frac{9599275547}{149299200}g^8 +\cdots
\end{eqnarray}
As in the previous section, we actually continue this expansion up to the order of $g^{40}$, then use $(20,20)$ Pad\'e to extract the information from the divergent series. The results are shown as lines in Fig.~\ref{fig:HH_Re}.

The results from HO expansions are shown as crosses in the same figure. Once again, we found the Pad\'e expansions fit the numerical results up to the crossing points.  

\begin{figure}[htbp]
\begin{center}
\includegraphics[width=0.75\textwidth]{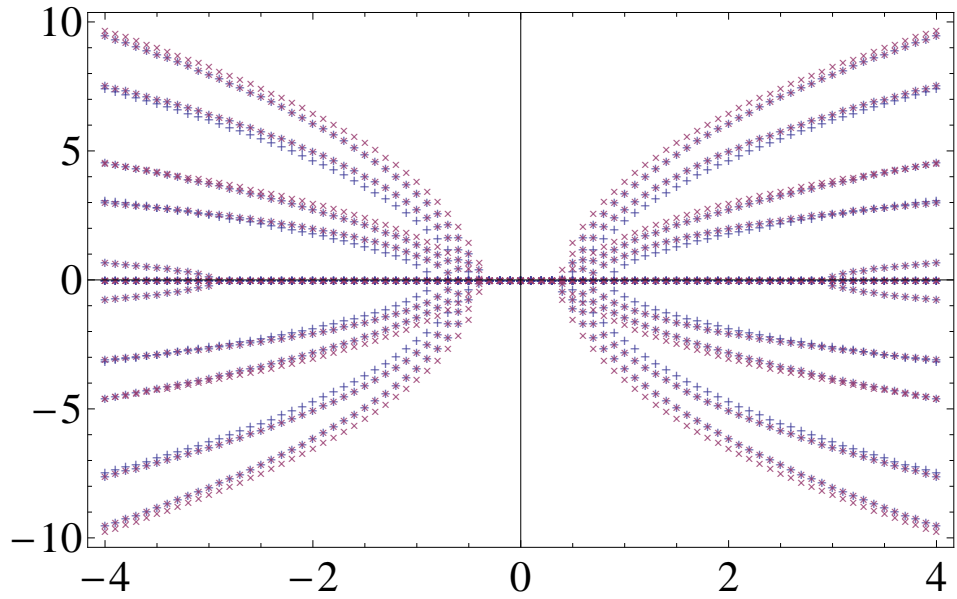}
\end{center}
\caption{\small
Imaginary parts of the eigenvalues as functions of the coupling constant $g$ with the complex H\'enon-Heiles potential. Only the results from HO expansions are shown. Even $y$-parity states are using ``$+$'' and odd $y$-parity states are using ``$\times$''.}
\label{fig:HH_Im}
\end{figure}

The ${\cal PT}$ symmetry is also broken in this Hamiltonian. Some eigenvalues cross and form complex conjugate pairs. The imaginary parts of the eigenvalues by HO expansions are shown in Fig.~\ref{fig:HH_Im}.

\section{Concluding remarks}
\label{sec:conclusion}

We concluded that both Hamiltonians have complex spectra. The origin of this broken ${\cal PT}$ symmetry is unclear. It could be due to the interaction between Hermitian and non-Hermitian potentials as in Ref.~\cite{BJ}. It could also be due to the level contraction instead of level repulsion in the non-Hermitian quantum mechanics. 

The results indicate that there is no critical value of the coupling constant to have reality in this degenerated setup. In the analogue with the $igxy$ interaction studied in Ref.~\cite{NanaPLA2}, it is possible that small enough coupling constant exhibits entire real spectrum in the non-degenerated setup. This would be confirmed by a future study.

\bigskip
{\bf Acknowledgment}
\smallskip

The author is very grateful to Professor C.M.~Bender and Professor G.V.~Dunne for much useful advice during the production of this paper.


\begin{thebibliography}{99}
\bibitem{BB}
C.M.~Bender and S.~Boettcher,
``{\it Real Spectra in Non-Hermitian Hamiltonians Having PT Symmetry},''
Phys.~Rev.~Lett.~{\bf 80}, 5243-5246 (1998)

\bibitem{Review}
C.M.~Bender,
``{\it Making Sense of Non-Hermitian Hamiltonians},''
Rep.~Prog.~Phys.~{\bf 70}, 947-1018 (2007) [arXiv:hep-th/0703096] 

\bibitem{BDMS}
C.M.~Bender, G.V.~Dunne, P.N.~Meisinger, and M.~\d{S}im\d{s}ek,
``{\it Quantum complex H\'enon-Heiles potentials},''
Phys.~Lett.~A {\bf 281}, 311-316 (2001)

\bibitem{NanaPLA1}
A.~Nanayakkara and C.~Abayaratne,
``{\it Semiclassical quantization of complex H\'enon-Heiles systems},''
Phys.~Lett.~A {\bf 303}, 243-248 (2002)

\bibitem{NanaPLA2}
A.~Nanayakkara,
``{\it Real eigenspectra in non-Hermitian multidimensional Hamiltonians},''
Phys.~Lett.~A {\bf 304}, 67-72 (2002)

\bibitem{NanaPLA3}
A.~Nanayakkara,
``{\it Comparison of quantal and classical behavior of ${\cal PT}$-symmetric systems at avoided crossings},''
Phys.~Lett.~A {\bf 334}, 144-153 (2005)

\bibitem{BTZPLA}
H.~B\'ila, M.~Tater, M.~Znojil,
``{\it Comment on: `Comparison of quantal and classical behavior of ${\cal PT}$-symmetric systems at avoided crossings' [Phys.~Lett.~A 334 (2005) 144]},''
Phys.~Lett.~A {\bf 351}, 452-456 (2006)

\bibitem{Mathematica}
M.~Trott, 
``Mathematica GuideBooks for Symbolics,'' 
Springer-Verlag, New York (2004)

\bibitem{BJ}
C.M.~Bender and H.F.~Jones,
``{\it Interactions of Hermitian and non-Hermitian Hamiltonians},''
J.~Phys.~A: Math.~Theor.~{\bf 41}, 244006 (2008) [arXiv:0709.3605]

\end{thebibliography}
\end{document}